\documentstyle[graphics]{mn}
\newcommand{\etal}{{\it et al.}~}

\newcommand{\bc}{\begin{center}}
\newcommand{\be}{\begin{equation}}
\newcommand{\ee}{\end{equation}}
\newcommand{\ec}{\end{center}}

\newcommand{\apm}{{\sc {APMSPH }}}
\newcommand{\hydra}{{\sc{HYDRA }}}

\newcommand{\cmbfast}{{\sc {CMBFAST }}}
\newcommand{\vpfit}{{\sc {VPFIT }}}
\newcommand{\lya}{Ly$\alpha$~}
\newcommand{\eg}{{\it e.g.~}}

\renewcommand{\H}{{\mbox{${\rm H{\sc i}~}$}}}
\newcommand{\Hp}{{\mbox{${\rm H{\sc ii}~}$}}}
\newcommand{\He}{{\mbox{${\rm He{\sc i}~}$}}}
\newcommand{\Hep}{{\mbox{${\rm He{\sc ii}~}$}}}
\newcommand{\Hepp}{{\mbox{${\rm He{\sc iii}~}$}}}
\newcommand{\h} {{\rm H{\sc i}}}

\newcommand{\ltsima}{\mbox{$\; \buildrel < \over \sim \;$}}
\def \simlt{\lower.5ex\hbox{\ltsima}}            
\def \gtsima{\mbox{$\; \buildrel > \over \sim \;$}}
\def \simgt{\lower.5ex\hbox{\gtsima}}            

\title[Cosmological dependencies of QSO \lya statistics]{Dependencies
of QSO \lya absorption line statistics on cosmological parameters}

\author[Theuns \etal]{Tom Theuns$^1$, Anthony Leonard$^2$, Joop Schaye$^1$ and
George Efstathiou$^1$\\
$^{\,1}$ Institute of Astronomy, Madingley Road, Cambridge CB3 0HA, UK\\
$^{\,2}$ Department of Physics, Astrophysics, University of Oxford, Keble Road, Oxford OX1 3RH, UK
}

\begin{document}

\maketitle

\begin{abstract}

We have performed high-resolution hydrodynamic simulations of the \lya
forest in a variety of popular cold dark matter dominated cosmologies,
including a low density and a vacuum-dominated model. The fluctuation
amplitude of these models is chosen to match the observed abundance of
galaxy clusters at low redshift. We assume that the intergalactic
medium is photoionized and photoheated by a uniform UV-background with
the required amplitude to give the observed mean hydrogen absorption.
We produce simulated spectra, analyze them by fitting Voigt profiles
and compare line statistics with those obtained from high resolution
observations. All models give column density distributions in good
agreement with observations.  However, the distribution of line widths
(the $b$-parameter distribution) reflects differences in the
temperature of the intergalactic medium between the models, with colder
models producing more narrow lines. All the models with a low baryon
density, $\Omega_b h^2=0.0125$, are too cold to produce a $b$-parameter
distribution in agreement with observations. Models with a higher
baryon density, $\Omega_b h^2=0.025$, are hotter and provide better
fits. Peculiar velocities contribute significantly to the line widths
in models with low matter density, and this improves the agreement with
observations further. We briefly discuss alternative mechanisms for
reconciling the simulations with the observed $b$-parameter
distributions.
\end{abstract}


\section{Introduction}
Neutral hydrogen in the intergalactic medium produces a forest of \lya
absorption lines blueward of the \lya emission line in quasar spectra
(Bahcall \& Salpeter 1965, Gunn \& Peterson 1965). Hydrodynamic
simulations of hierarchical structure formation in a cold dark matter
(CDM) dominated universe have been shown to be remarkably successful in
reproducing a variety of properties of the observed forest (Cen \etal
1994, Zhang, Anninos \& Norman 1995, Miralda-Escud\'e \etal 1996,
Hernquist \etal 1996, Wadsley \& Bond 1996, Zhang \etal 1997, Theuns
\etal 1998ab, Dav\'e \etal 1998, Bryan \etal 1998). The simulations
show that the weaker \lya lines (neutral hydrogen column density
$N_\H\le 10^{14}$ cm$^{-2}$) are predominantly produced in the
filamentary and sheet-like structures that form naturally in this
structure formation scenario. Higher column density lines occur where
the line of sight intersects a halo.

These models assume that the intergalactic medium (IGM) is
photoionized and photoheated by UV-light from quasars. The
characteristic break in the rate of evolution of the number of lines
below a redshift $\sim 1.7$, as observed by the {\em Hubble Space
Telescope}, can then be explained by the decrease in the intensity of
this ionizing background, itself a consequence of the rapid decline in
quasar numbers towards lower redshifts (Theuns, Leonard \& Efstathiou
1998, Dav\'e \etal 1998).

While the first simulations showed good agreement with the observed
line statistics, more detailed studies at higher numerical resolution
in a standard CDM universe produced a larger fraction of narrow lines
than are observed (Theuns \etal 1998b, Bryan \etal 1998). Theuns \etal
suggested that an increase in temperature of the IGM might broaden the
absorption lines sufficiently to improve the agreement with
observations. A higher gas temperature leads to more thermal broadening
and in addition increases the Jeans mass. At low redshifts, the
temperature of the gas responsible for the majority of \lya lines is
determined by the balance between adiabatic cooling and
photoheating. This causes the gas temperature to be a function of
density, and this \lq equation of state\rq~ is well approximated by a
power law, $T=T_0 (1+\delta)^{\gamma-1}$, where $\delta$ is the gas
overdensity (Hui \& Gnedin 1997). $T_0$ can be made higher by
increasing the photoheating rate, or by heating the gas for longer by
increasing the age of the universe. The first can be achieved by
increasing the baryon density $\Omega_b h^2$, the latter by decreasing
the matter density $\Omega_m$. Using some simplifying assumptions, Hui
and Gnedin (1997) obtain the scaling
\begin{equation}
T_0\propto \Large[\Omega_b h^2/ (\Omega_m h^2)^{1/2}\Large]^{1/1.7}\,.
\label{eq:eos}
\end{equation}
Increasing $\Omega_b h^2$ by a factor 2 and using $\Omega_m=0.3$
instead of 1 increases $T_0$ by a factor $\ge 2$, which might be
sufficient to obtain agreement with observations.

In this {\em Letter} we use high resolution hydrodynamic simulations to
investigate the dependence of the $b$-parameter distribution on
cosmological parameters, for a given reionization history. This
complements the work of Haehnelt \& Steinmetz (1998), who quantified
the dependence of the $b$-parameter distribution on the reionization
history, for a critical density Universe.

\section{Simulation}
\label{sect:simulation}
\begin{table}
 \centering \begin{minipage}{70mm} \caption{Models simulated}
 \begin{tabular}{@{}llllllll@{}} 
  & Name & $\Omega_m$ & $\Omega_\Lambda$ & $\sigma_8$ & $\Omega_b h^2$ & $h$ & $\Gamma_\h$ \\[5pt]
1 & S    & 1      &   0    & 0.50 & 0.0125  & 0.5  & HM/2 \\
2 & O   	& 0.3    &   0    & 0.85 &	0.0125  & 0.65 & HM/2 \\
3 & L    & 0.3    &   0.7  & 0.90 &	0.0125  & 0.65 & HM/2 \\
4 & Sb   & 1      &   0    & 0.50 &	0.025   & 0.5  & HM   \\
5 & Ob   & 0.3    &   0    & 0.85 &	0.025   & 0.65 & HM   \\
6 & Lb   & 0.3    &   0.7  & 0.90 &	0.025   & 0.65 & HM   \\

\label{table:runs}
\end{tabular}
\end{minipage}
\end{table}
\begin{figure}
\setlength{\unitlength}{1cm}
\centering
\begin{picture}(7,14)
\put(-4, -8){\includegraphics{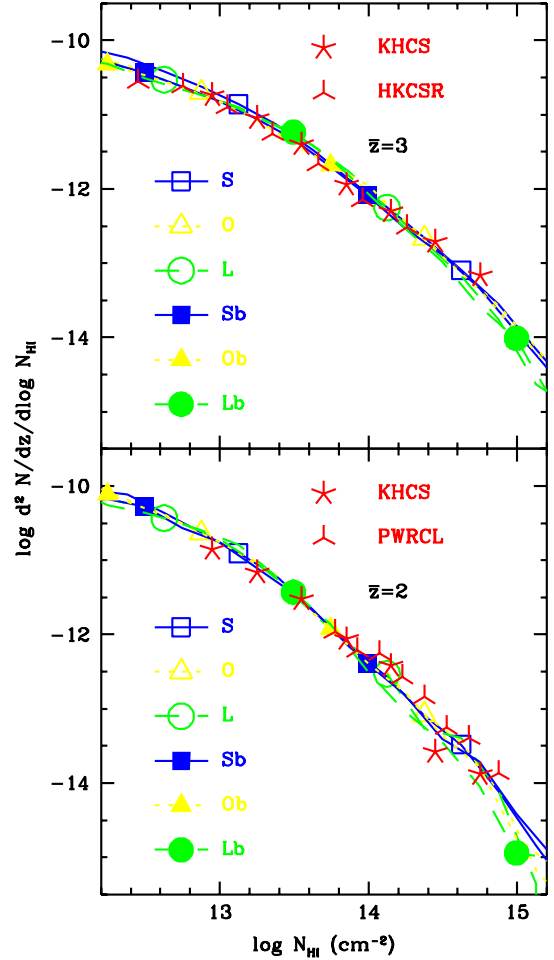}}
\end{picture}
\caption{Comparison between observed and simulated column density
distributions at mean redshifts of 3 (top panel) and 2 (bottom
panel). The data in the top panel are from Kim \etal (1997, hereafter
KHCS, $\bar z=3.35$) and Hu \etal (1995, hereafter HKCSR, $\bar z=2.9$)
and in the bottom panel from KHCS ($\bar z=2.3$) and from Petitjean
\etal (1993, hereafter PWRCL, $\bar z=2.0$). The different models are
indicated in the panel.}
\label{fig:ddf}
\end{figure}

\begin{figure*}
\setlength{\unitlength}{1cm}
\centering
\begin{picture}(21,14)
\put(-4, -8.){\includegraphics{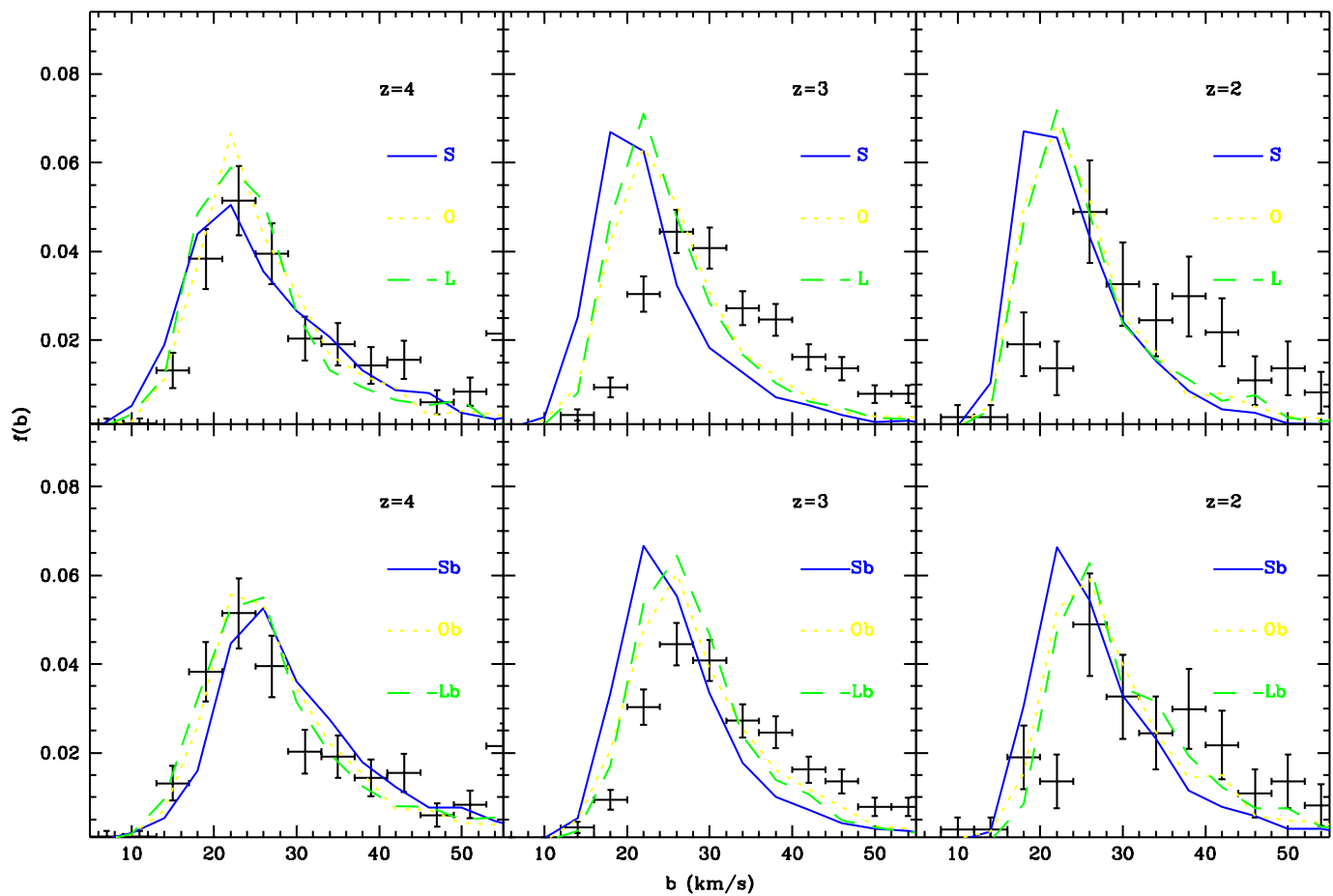}}
\end{picture}
\caption{Simulated $b$-parameter distribution for models with low
baryon density (top panels) and models with high baryon density (bottom
panels) at redshifts 4, 3 and 2 (left to right). Superimposed with
error bars are the observed data for $z=4$ from Lu \etal (1996), HKCSR
($z=3$) and KHCS ($z=2$). We have only included lines with column
density $N_\h>10^{13}$ cm$^{-2}$ that have an estimated error in $b$,
$b_{\em err}<0.25 b$, for both simulations and data.}
\label{fig:bpars}
\end{figure*}

We have simulated six different cosmological models, characterized by
their total matter density $\Omega_m$, the value of the cosmological
constant $\Omega_\Lambda$, the rms of mass fluctuations in spheres of
8$h^{-1}$ Mpc, $\sigma_8$, the baryon density $\Omega_b h^2$ and the
present day value of the Hubble constant, $H_0=100 h $ km
s$^{-1}$Mpc$^{-1}$. The parameters for these models are summarized in
Table~\ref{table:runs}. Note that the Sb, Ob and Lb models have a
baryon density slightly higher than the value $\Omega_b h^2=0.0193\pm
0.0014$, advocated by Burles \& Tytler (1998) from measurements of the
deuterium abundance in quasar spectra. Rauch \etal (1997) required a
similarly high value $\Omega_b h^2\ge 0.017$ in order to make the
measured flux decrement of the \lya-forest consistent with the observed
intensity of ionizing photons from quasars.

We model the evolution of a periodic, cubical region of the universe of
comoving size 2.5$h^{-1}$Mpc. The code used is based on a hierarchical
P3M implementation (Couchman 1991) for gravity and smoothed particle
hydrodynamics (SPH, Lucy 1977, Gingold \& Monaghan 1977, see \eg
Monaghan 1992 for a review) for hydrodynamics. It is a hybrid between
the \hydra code of Couchman \etal (1995) and the \apm code, which were
described in detail in Theuns \etal (1998b). In particular, we use the
\apm method for consistently computing SPH densities for particles in
underdense regions. These simulations use $64^3$ particles of each
species, so the SPH particle masses are $1.65\times 10^6\,(\Omega_b
h^2/0.0125) (h/0.5)^{-3} M_\odot$ and the CDM particles are more
massive by a factor $\Omega_{\rm CDM}/\Omega_b$. This resolution is
sufficient to simulate line widths reliably (Theuns \etal 1998b; note
that numerical convergence will be even better in the hotter
simulations). The simulations of Bryan \etal (1998) indicate that the
absence of long waves will produce a small, but for our purposes
unimportant, underestimate of the widths of the absorption
profiles. All our simulations were run with the same initial phases to
minimize cosmic variance when comparing the different models.

We assume that the IGM is ionized and photoheated by an imposed uniform
background of UV-photons that originates from quasars, as computed by
Haardt \& Madau (1996). This flux is redshift dependent, due to the
evolution of the quasar luminosity function. Haardt \& Madau (1996)
give two fits to the hydrogen ionization rate $\Gamma_\h$ as a function
of redshift. We have used their $q_0=0.5$ fit for the critical density
models and their $q_0=0.1$ fit for the open ones. This amplitude of the
flux is indicated as \lq HM\rq~ in the $\Gamma_\h$ column of
Table~\ref{table:runs}. In addition, for the low $\Omega_b h^2$ models,
we have divided the ionizing flux by two, indicated as \lq
HM/2\rq~. The dependence on $q_0$ reflects small differences in the
completeness corrections of the observed quasar luminosity functions
and is relatively unimportant for the simulations described here. The
detailed expressions for the heating and cooling rates as a function of
temperature and ionizing flux are taken from Cen (1992) with some minor
modifications (Theuns \etal 1998b). Our analysis differs slightly from
that in Theuns \etal in that we do not impose thermal equilibrium but
solve the rate equations to track the abundances of $\H$, $\Hp$ and
$\He$, $\Hep$ and $\Hepp$. We assume a helium abundance of $Y=0.24$ by
mass.

We have used \cmbfast (Seljak \& Zaldarriaga 1996) to compute the
linear matter transfer function for each model. The amplitude of the
transfer function is normalized to the observed abundance of galaxy
clusters at $z=0$ using the fits $\sigma_8=0.52
\Omega_m^{-0.46+0.1\Omega_m}$ for $\Omega_\Lambda=0$ and $\sigma_8=0.52
\Omega_m^{-0.52+0.13\Omega_m}$ for $\Omega_\Lambda=1-\Omega_m$, as
computed by Eke, Cole \& Frenk (1996).

At several output times we compute simulated spectra along lines of
sight through the simulation box. Each spectrum is convolved with a
Gaussian with full width at half maximum of FWHM = 8 km s$^{-1}$, then
re-sampled onto pixels of width 3 km s$^{-1}$ to mimic the instrumental
profile and characteristics of the HIRES spectrograph on the Keck
telescope. We rescale the background flux in the analysis stage such
that the mean effective optical depth at a given redshift in all models
is the same as for the Ob model . The latter model has a mean
absorption in good agreement with observations (Ob has $\bar\tau_{\rm
eff}=0.93$, 0.33 and 0.14 at $z$=4, 3 and 2). Finally, we add to the
flux in every pixel a Gaussian random signal with zero mean and
standard deviation $\sigma=0.02$ to mimic noise. We fit the continuum
of the spectra using the method described in Theuns \etal (1998b). The
absorption features in these mock observations are then fitted with
Voigt profiles using an automated version of \vpfit (Carswell \etal
1987).

\section{Results and discussion}
A comparison between the observed and simulated column density
distribution functions (CDDFs) is shown in Fig.~\ref{fig:ddf}. The
CDDFs for the different models are almost indistinguishable and in
addition are very similar to the observed distribution at both
redshifts 2 and 3. In principle, the slope of the CDDF is sensitive to
the amplitude of the mass fluctuations on small scales (\eg Gnedin
1998). Figure~\ref{fig:ddf} then shows that demanding consistency of
the model with the observed abundance of galaxy clusters at $z=0$
automatically guarantees good agreement with the observed CDDF, for the
CDM dominated cosmological models examined here.

The distribution of line-widths for the different models is compared
with observation in Fig.~\ref{fig:bpars}. We have selected those lines with
column density $N_\h>10^{13}$ cm$^{-2}$ that have an estimated error
from \vpfit $b_{\em err}<0.25 b$. For the data we have applied the same
column density cut, and used the published uncertainties in $b$ for
each fitted \lya-line, where available.

The models have different temperature-density relations at $z=4$, yet
all of them provide $b$-parameter distributions which are in good
agreement with the observed distribution at this redshift. However, at
$z=3$ there is a significant overproduction of narrow lines in all
models with $\Omega_b h^2=0.0125$, compared to the data. The models
with the higher value $\Omega_b h^2=0.025$ agree much better with the
observed $b$-distribution, although the critical density Sb model still
peaks at lower values than the observed distribution. The open models
Ob and Lb, on the other hand, produce $b$-parameter distributions which
fit the observed distribution reasonably well. At redshift $z=2$, the
quality of the data is lower since the Keck spectrograph becomes less
efficient in the blue. The high baryon models produce $b$-parameter
distributions which fit the observed distribution at $z=2$ better than
the low $\Omega_b h^2$ models. Again, the critical density model peaks
at significantly lower $b$ than the observed distribution.

\begin{figure*}
\setlength{\unitlength}{1cm}
\centering
\begin{picture}(21,7)
\put(-4., -8){\includegraphics{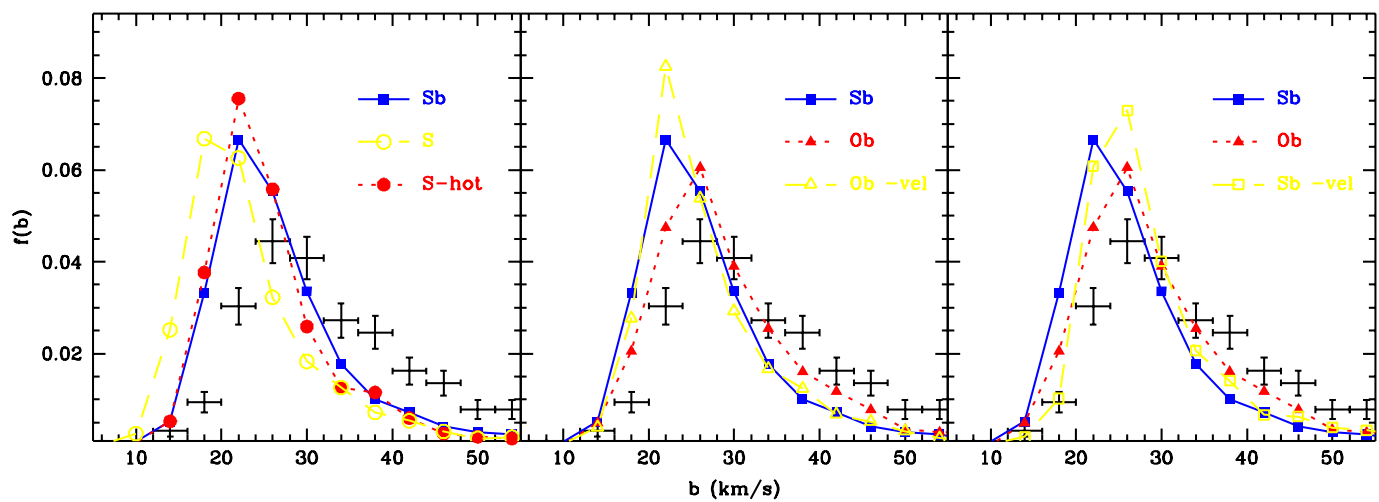}}
\end{picture}
\caption{Simulated and observed $b$-parameter distributions at
$z=3$. Models Sb, S and Ob are taken from Fig.\ref{fig:bpars}. The left
panel illustrates the effect of the equation of state: model S-hot is
obtained from model S by imposing the equation of state from the hotter
model Ob. Middle and right panels illustrate the effect of peculiar
velocities: models Ob-vel and Sb-vel are obtained from Ob and Sb by
neglecting peculiar velocities. Ob and Sb are shown for comparison.
Error bars denote the data from HKCSR.}
\label{fig:bparcosmo}
\end{figure*}

We illustrate the main reasons for the cosmology dependence of the
$b$-parameter distribution in Fig.~\ref{fig:bparcosmo}. We have
recomputed simulated spectra for model S after imposing the equation of
state of the significantly hotter model Ob. We did this by increasing
the temperature of SPH particles when they were cooler than the
temperature given by the required equation of state. We then rescaled
the flux so that the mean decrement was again the same as for the Ob
model. Comparing this new model S-hot with S and Sb, we see that the
main reason for the broader lines in Sb as compared to S is the higher
temperature of the IGM in model Sb. However, the $b$-distribution of
model S-hot still peaks at lower values than that of model Ob. The
reason for the remaining differences is the contribution of peculiar
velocities to the line-width for model Ob, as shown in
Fig.~\ref{fig:bparcosmo}. Model \lq Ob-vel\rq~ is identical to model
Ob, except that we have put all peculiar velocities to zero. This
increases the fraction of narrow lines and the $b$-distribution of this
new model is almost identical to that of models Sb and S-hot. The right
panel in Fig.~\ref{fig:bparcosmo} shows that neglecting peculiar
velocities in model Sb {\em decreases} the fraction of narrow lines for
this cosmology, as shown by model Sb-vel. Consequently, the open model
Ob has broader lines than the critical model Sb because of the
different way that peculiar velocities contribute to line broadening
between the two models. The effects of peculiar velocity are similar in
models Lb (not shown) and Ob.

In summary, we have investigated the properties of the simulated
\lya-forest in three popular cosmologies (critical, open and
vacuum-dominated) and compared simulated spectra analyzed in terms of
Voigt profiles with observations. When the power spectrum is normalized
to the abundance of galaxy clusters and the amplitude of the ionizing
background is normalized to match the observed mean absorption, then
all investigated models produce column density distributions which are
well matched to the observations, at $z=2$ and 3. However, the
distribution of $b$-parameters is sensitive to the baryon density as
well as to the matter density of the universe. Models with a higher
baryon density $\Omega_b h^2$ are hotter and hence the increased
thermal broadening and Jeans smoothing shift the $b$-parameter
distribution towards larger $b$ values, providing a better fit to the
observations. Peculiar velocities contribute significantly to the
broadening of absorption lines with column density $N_\h \sim 10^{13}$
cm$^{-2}$, in low matter density models. This broadening improves the
agreement with observations further. In contrast, peculiar velocities
narrow the lines in a critical density universe. The $b$-distributions
for our open and vacuum-dominated models with $\Omega_m=0.3$ are almost
identical. Of all our simulated models, the one that fits best has
$\Omega_b h^2=0.025$ and $\Omega_m=0.3$ and produces significantly more
broad lines than the other models, yet there is a hint that the
observed distribution has an even larger fraction of broad lines. It is
not clear whether the present data are sufficiently reliable that this
is a real discrepancy. For example, some of the remaining differences
may be due to differences in the Voigt profile fitting procedures
between \vpfit and HKCSR or due to uncertainties in the continuum
fitting (Rauch \etal 1993), although it seems unlikely that these
effects would be large enough to reconcile our model with the data. If
the discrepancy is real, then it is possible that feedback from \eg
star formation, not included in this investigation, might contribute to
broadening the lines. Alternatively, a larger contribution from helium
heating than assumed in our model might be required to boost the
temperature further.

\section*{Acknowledgments}
AL thanks PPARC for the award of a research studentship, JS thanks the
Isaac Newton Trust, St. John's College and PPARC for support and GE
thanks PPARC for the award of a senior fellowship. We thank M.
Haehnelt for many stimulating discussions and R. Carswell for helping
us with \vpfit.

{}
\end{document}